\documentclass[12pt]{article}

\usepackage{paper2e}
\usepackage{mydefs2e}
\usepackage{equations}

\renewcommand{\d}{\partial}

\begin{document}

\begin{titlepage}
\preprint{TU-913}

\title{Partially Composite Higgs 
in Supersymmetry}

\author{Ryuichiro Kitano$^{1}\!$,\ \ 
Markus A. Luty$^{2,3}\!$,\ \ 
Yuichiro Nakai$^{1}$}

\address{$^1\,$%
Department of Physics, Tohoku University\\
Sendai Japan}

\address{$^2\,$%
Institut f\"ur Theoretische Physik der Universit\"at Heidelberg\\
Hiedelberg, Germany}

\address{$^3\,$%
Physics Department, University of California Davis\\
Davis, California USA} 

\vspace*{1.0cm}

\begin{abstract}
We propose a framework for natural breaking of electroweak
symmetry in supersymmetric models, where elementary Higgs
fields are semi-perturbatively coupled to a strong superconformal sector.
The Higgs VEVs break conformal symmetry in the strong sector
at the TeV scale, and the strong sector in turn gives
important contributions to the Higgs potential,
giving rise to a kind of Higgs bootstrap.
A Higgs with mass $125\GeV$ can be accommodated without any
fine tuning.
A Higgsino mass of order the Higgs mass is also dynamically
generated in these models.
The masses in the strong sector generically violate custodial symmetry,
and a good precision electroweak fit requires tuning of order
$\sim 10\%$.
The strong sector has an approximately supersymmetric spectrum of
hadrons at the TeV scale that can be observed by looking for
a peak in the $WZ$ invariant mass distribution, as well as
final states containing multiple $W$, $Z$, and Higgs bosons.
The models also generically predict large corrections 
(either enhancement or suppression) to the
$h \to \ga\ga$ width.
\end{abstract}

\end{titlepage}

\section{Introduction}
Supersymmetry (SUSY) remains a compelling framework for addressing
the naturalness problem of electroweak symmetry breaking
\cite{Martin:1997ns}.
The recent evidence for a 125~GeV Higgs boson at the LHC~\cite{LHCHiggs}
motivates us to ask whether such a Higgs mass is compatible
with naturalness in the context of supersymmetry.

A 125~GeV Higgs requires significant
tuning in the minimal supersymmetric standard model (MSSM).
The reason is that we need a significant radiative correction 
to the Higgs quartic from top and stops, the particles most strongly
coupled to the Higgs:
\beq[Higgsquartictop]
\De \la_H \sim \frac{y_t^4 N_c}{16\pi^2} \ln \frac{m_{\tilde t}}{m_t},
\eeq
where $N_c = 3$.
The stop mass is generally lighter than the other squark masses
due to renormalization group effects, so this tends to push
much or all of the superpartner spectrum is out of reach of the
LHC.
However, there is a good reason to think that this is not how nature
works: such models are highly fine-tuned.
The reason is that the large stop mass also generates
a large quadratic term in the Higgs potential
that must be tuned away:
\beq[Higgsquadratictop]
\De m_H^2 \sim \frac{y_t^2 N_c}{16\pi^2} m_{\tilde t}^2 
\ln \frac{M^2}{m_{\tilde{t}}^2},
\eeq
where $M$ is a UV mass scale where the stop mass is generated.
We note that such quadratic dependence of the Higgs mass parameter
on large mass scales is precisely the naturalness problem that SUSY is  
supposed to address.
Numerically, the tuning is of order a percent, even if the logarithm
in \Eq{Higgsquadratictop} is not large.
It is therefore well-motivated to consider
possible mechanisms to reduce this tuning and study
their experimental implications.

In order to generate a large quartic without fine-tuning,
what is required are Higgs interactions that are stronger than 
typical perturbative interactions.
We can see this from \Eqs{Higgsquartictop} and \eq{Higgsquadratictop}:
if we could increase $y_t$, then $\De\la_H$ increases as $y_t^4$
while $\De m_H^2$ only increases as $y_t^2$.
In this paper, we consider the possibility that the Higgs is coupled
to a strong sector, so the light Higgs is partially composite.
This arises naturally in a supersymmetric model if the Higgs is coupled to
a strong superconformal sector via operators
\beq[HiggsCFTint]
W = \la_u H_u \scr{O}_d + \la_d H_d \scr{O}_u,
\eeq
where $\scr{O}_{u,d}$ are operators in the conformal sector with
the electroweak quantum numbers of $H_{u,d}$.
The dimension $d$ of the operators $\scr{O}_{u,d}$ can be smaller
than 2, so that the couplings above are relevant.
This means that there is no UV problem with the interactions in
\Eq{HiggsCFTint} being stronger than perturbative interactions at the 
TeV scale.
This theory can therefore easily accommodate heavy
Higgs masses without fine tuning or UV problems.
This is the primary motivation for considering such models.

On the other hand, the fact that these couplings are relevant
introduces a new coincidence problem similar to the $\mu$-problem
of the MSSM, namely why the scale of the SUSY-preserving terms
\Eq{HiggsCFTint} is near the weak scale.
We will discuss a generalization of the Giudice-Masiero mechanism
that can address this problem.

\Ref{SCTC} studied the phenomenology of such models in the case where
SUSY breaking at the TeV scale triggers confinement and dynamical
electroweak symmetry breaking in the strong sector, which in turn
induces VEVs for the elementary Higgs fields.
(Such models were called ``superconformal technicolor.'')
In this paper, we consider a different limit where the strong
sector does not break electroweak symmetry in the limit 
$\la_{u,d} \to 0$.
We assume that the dominant contribution to conformal symmetry breaking
in the strong sector comes from the Higgs VEVs.
The strong sector then gives important contributions to
the effective potential of the elementary Higgs, so
the model is a kind of Higgs bootstrap.
These models have a very different phenomenology from superconformal
technicolor models, as we will see below.
Holographic 5D models with Higgs couplings of form \Eq{HiggsCFTint}
were studied in \Ref{PG}, assuming that SUSY is broken in the strong
sector near the TeV scale.
Models similar to ours were studied in \Ref{Vafa} with particular attention
to the case where the operators $\scr{O}_{u,d}$ have dimension near 2.
See also \Ref{Fukushima:2010pm} for a semi-perturbative model
which appears as a dual description of a strongly coupled theory.
\Ref{Yanagida}, which appeared while this paper was being completed,
also considers semi-perturbative conformal sectors, and is closely
related to the present work.

In the presence of the interactions \Eq{HiggsCFTint},
electroweak symmetry breaking in
the weakly-coupled sector induces breaking of conformal and electroweak
symmetry in the strong sector at the TeV scale, in addition to explicit
conformal symmetry breaking by soft SUSY breaking.
We also consider the next-to-minimal supersymmetric standard model (NMSSM),
where the VEV of the singlet is an important contribution to conformal
symmetry breaking in the strong sector.  
These models eliminate a potential problem in superconformal technicolor,
namely the presence of unstable potential in the strong sector \cite{SCTC}.  
In the absence of any tuning, we will see that the
electroweak symmetry breaking masses in the strong sector are of the
same size as the electroweak preserving masses, and the precision
electroweak corrections are quite large.  
However, we will show that $\sim 10\%$ tuning is sufficient to
reduce the precision electroweak corrections, so the model is much
less tuned than the MSSM.

This paper is organized as follows.
In \S 2, we describe specific models to set the stage for the
more general discussion that follows.
We consider the case where the dominant breaking
of conformal symmetry in the strong sector comes from 
electroweak doublet or singlet Higgs fields.
In \S 3, we give a general discussion of this class of models
and estimate the corrections to the effective potential for the elementary
Higgs fields.
In \S 4, we give estimates for
the precision electroweak corrections.
In \S 5, we consider the phenomenology, 
and \S 6 gives our conclusions.
In appendix, we discuss the contribution of soft SUSY breaking
terms in the strong sector to the Higgs potential.

\section{Models
\label{sec:models}}
We begin by presenting some specific models that illustrate the general
ideas.

\subsection{Models with Custodial Symmetry
\label{sec:modelscustodial}}
The minimal model is based on an $SU(2)$ strong gauge group
with $4$ flavors (8 fundamentals).
This theory is in the middle of the conformal window,
and has no known weakly coupled description.
The lowest-dimension chiral primary field (a meson) has
dimension $\frac 32$.
The fact that this is not close to the free-field dimension 1
is an indication that this is a truly strongly-coupled theory.

The gauge group of the model is
\beq
SU(2)_S \times SU(2)_L \times SU(2)_R \times U(1)_{B-L},
\eeq
where the electroweak gauge group is embedded into
$SU(2)_L \times SU(2)_R \times U(1)_{B-L}$
in the standard way,
\ie $SU(2)_L = SU(2)_W$ and $Y = T_{3R} + B - L$.
The MSSM Higgs fields $H_{u,d}$ are therefore contained in
the field
\beq
H \sim (1, 2, 2)_0.
\eeq
This embedding of electroweak gauge group allows a natural
custodial symmetry to act on the fields of the strong sector,
namely the diagonal subgroup of $SU(2)_L \times SU(2)_R$.
On the other hand, multiplets of $SU(2)_L \times SU(2)_R$
are not complete GUT multiplets, so automatic gauge coupling
constant unification is lost in these models.

There are various possibilities for the fields.
One simple possibility is
\beq
\bal
\Psi_i &\sim (2, 2, 1)_b,
\\
\tilde\Psi_i &\sim (2, 1, 2)_{-b},
\eal
\eeq
where $i = 1, 2$ and $b$ is the $B - L$ charge.
In the custodial symmetry limit, we can write the Higgs coupling
to this sector as
\beq[Higgscoupminimalmodel]
W = \la_{ij} H \Psi_i \tilde{\Psi}_j.
\eeq
The ``meson'' fields $\Psi \tilde{\Psi}$ have scaling dimension
$\frac 32$, so the couplings $\la$ have dimension $+\frac 12$,
\ie\ they are relevant.

The Higgs VEVs break the conformal symmetry in the strong
sector, giving positive supersymmetric masses to all matter fields
in the strong sector.
The $SU(2)_S$ gauge fields (and gauginos) are massless
and confine, generating a dynamical superpotential
\beq
W_{\rm dyn} \sim \la^2 H_u H_d.
\eeq
The model therefore generates a $\mu$ term dynamically.

There are also contributions to the Higgs potential arising
from SUSY breaking scalar and gaugino masses in the strong sector.
However, if SUSY breaking is transmitted to the strong sector at a
high scale, large anomalous dimensions suppress the gaugino mass
and universal scalar masses \cite{SUSYCFTsoft,NDAN2}.
The result is that the soft masses at low energies must satisfy
\beq
\sum_i (m_i^2 + \tilde{m}_i^2) = 0
\eeq
where $m^2_i$ ($\tilde{m}_i^2$) are the scalar masses for field $\Psi_i$
($\tilde\Psi_i$).
Such mass terms generally destabilize the vacuum of the strong
sector \cite{SCTC}, so we will assume that the strong sector masses
generated by the Higgs VEVs dominate.
These always give positive scalar masses, and there is no stability
problem in these theories.

Even though SUSY breaking is subleading in the strong sector it
can give important contributions to the Higgs potential,
which in turn determines the scale $\La$ of the strong sector.
This is therefore a kind of ``Higgs bootstrap.''

To see whether this model is realistic, we need to estimate
the size of these effects.
Because the fixed point gauge coupling is strong at all
scales, we expect that there is no hierarchy between the
confinement scale $\La$ and the scale of the masses induced by
the VEVs. 
We can therefore estimate the terms in the effective Lagrangian
using \naive\ dimensional analysis (NDA) \cite{NDA}.%
\footnote{%
Exact results for $\scr{N} = 2$ theories
suggest that NDA is accurate in SUSY theories \cite{NDAN2}.}
This will be done in \S\ref{sec:Higgspotential} below,
after we have discussed several additional models
to illustrate the range of possibilities.
We will see that the contributions to the Higgs potential
can be large enough to get a Higgs mass of $125\GeV$ without
tuning, but the model requires tuning of order $10\%$ to satsify
precision electroweak constraints.
This situation is more general than this specific model,
and we will present a general analysis in \S\ref{sec:PEW}
below.

A variation of this model is to allow electroweak preserving masses
for the strong sector.
These can be generated by adding a singlet Higgs field $S$
to the theory and writing the new superpotential terms
\beq
\De W \sim S \Psi \Psi + S \tilde{\Psi}\tilde{\Psi}.
\eeq
This is allowed only if we choose the $B - L$ charge $b$ to vanish.
We can have electroweak preserving masses without introducing a singlet
by adding ``$\mu$ terms'' to the superpotential, \ie\
$\De W \sim  \Psi \Psi + \tilde{\Psi}\tilde{\Psi}$.
However, such terms have a different dimension from the Higgs
couplings \Eq{Higgscoupminimalmodel}, 
and therefore have no reason to be of the same order.

The importance of adding electroweak-preserving masses is that the 
corrections to precision electroweak observables from the strong sector 
can be  reduced if the electroweak-preserving masses are larger than
the electroweak-breaking ones.
We will see that this requires parametrically larger tuning
than tuning toward a custodial symmetry limit, but it may work.
Again, we will give a general analysis after we have presented
several models.

Yet another variation is to have some of the strong fields be
electroweak singlets.
These can get a mass from coupling to $S$.
This can somewhat reduce the precision electroweak corrections,
and also the corrections to $h \to \ga\ga$, which we discuss
in \S\ref{sec:hgg} below.

\subsection{Models with Unification}
It is natural to consider models where the strong fields come in
complete $SU(5)$ multiplets, so that the theory unifies.
A simple example of such a model is based on a strong $SU(3)$
gauge group with 6 flavors.
This is again in the middle of the conformal window, and is 
a strongly-coupled theory.

The gauge group is
\beq
SU(3)_S \times SU(5)_{\rm SM},
\eeq
where the standard model gauge group is embedded in
$SU(5)_{\rm SM}$ in the standard way.
The matter fields transform as
\beq
\bal
\Psi &\sim (3, 5),
\\
\tilde\Psi &\sim (\bar{3}, \bar{5}),
\\
\Si &\sim (3, 1),
\\
\tilde\Si &\sim (\bar{3}, 1).
\eal\eeq
The $\bf 5$ and $\bar{\bf 5}$ fields decompose into 
doublet and triplet fields
\beq
\Psi = (D, T),
\qquad
\tilde\Psi = (\tilde{D}, \tilde{T}),
\eeq
and we can write superpotential couplings
between the strong fields and Higgs fields as
\beq
W = \la_u H_u \tilde{D} \Si + \la_d H_d D \tilde\Si
+ \la_\Si S \tilde\Si \Si
+ \la_D S \tilde{D} D + \la_T S \tilde{T} T.
\eeq
Note that we need a singlet field $S$ to give masses to
all the strong fields that are naturally the same size
as those generated by Higgs VEVs.

One difficulty with this model is that the electroweak breaking
masses in the strong sector cannot preserve custodial symmetry.
This is because the electrically neutral component of the doublets
can get an electroweak-breaking mass with the
$\Si$ and $\tilde\Si$ fields, but there is no partner for the
charged component of the doublets.
As we will see below,
these models require parametrically more tuning to satisfy
precision electroweak constraints, but may still be viable.

\section{Partially Composite Higgs Potential
\label{sec:Higgspotential}}
In this section we give a general discussion of 
the contributions to the Higgs 
potential from the strong sector.
Our discussion is valid for a general strong sector,
but it is helpful to have the examples of the previous section
in mind to understand the discussion.

We write the superpotential coupling as
\beq[CFTHiggsint]
W = \ka_u^{2-d} H_u \scr{O}_d + \ka_d^{2-d} H_d \scr{O}_u,
\eeq
where $d$ is the dimension of $\scr{O}_{u,d}$, so that
$\ka$ has dimensions of mass.
We must have $d > 1$ by unitarity.
We assume $d < 2$, so that these couplings are relevant,
and normalize the operators so that the theory gets strong
at the scale $\ka$.
The case $d \simeq 2$ is special, since the couplings
are nearly dimensionless \cite{Vafa}.
We will comment on this case below, but will
focus mainly on the case where $2 - d$ is not a small
parameter.
The strength of the couplings is then measured by
\beq[epsilonkappa]
\ep_{u,d} \sim \left( \frac{\ka_{u,d}}{\La} \right)^{2-d}
\eeq
where $\La$ is the scale of conformal symmetry breaking
in the strong sector (see below).
These interactions cause the Higgs fields $H_{u,d}$ to mix
with the strong conformal sector, which can give important
contributions to the Higgs potential as we will see below.

$\La$ is determined by conformal symmetry breaking mass terms
in the strong sector.
(By mass terms, we mean relevant terms in the Lagrangian involving
only strong sector fields.)
We will focus on the possibility that the mass terms in the
strong sector induced by the Higgs VEVs via the coupling
\Eq{CFTHiggsint} are the largest breaking of conformal symmetry
and trigger the exit from the CFT fixed point.
Because the Higgs VEVs preserve SUSY, the 
strong sector is approximately supersymmetric.
The strong sector then in turn gives important contributions
to the Higgs potential, this is a kind of Higgs bootstrap.

It is also possible to have conformal symmetry breaking
that does not violate electroweak symmetry from
explicit $\mu$-like terms, or from the VEV of a perturbative
singlet field.
There is an interesting class of models where this is the
dominant source of conformal symmetry breaking in the strong
sector.
Again, the strong sector is naturally approximately supersymmetric
in this case.

Finally, there is soft SUSY breaking in the strong sector.
As discussed in \S\ref{sec:modelscustodial}, 
if this is the dominant source of conformal symmetry 
breaking in the strong sector, then there are difficulties
with vacuum stability.
These can be overcome with some additional structure \cite{SCTC}
but we will focus on the case where the dominant
source of conformal symmetry breaking is supersymmetric.

\subsection{Higgs Bootstrap
\label{sec:Higgsbootstrap}}
We first consider the situation where electroweak breaking Higgs
VEVs trigger the exit from the CFT.
This occurs in the first model of \S\ref{sec:modelscustodial} above,
but may also occur in models with singlet fields if their VEVs
are subdominant.
In this case we have
\beq
\ep \sim \frac{\La}{4\pi v},
\label{eq:epsdef}
\eeq
where $\ep \sim \ep_{u,d}$ and we assume $v_u \sim v_d$.
We can understand this from the fact that in the double limit 
$\ep \sim 1$ and $4\pi v \sim \La$ both the relevant interactions
and the Higgs mass contributions are strong at the scale $\La$.

Note that the Higgs VEV is a SUSY-preserving mass in the strong
sector.
We therefore first consider the dynamics of the strong sector
in the SUSY limit.
The scale of strong dynamics is determined by the Higgs VEV,
and can be parameterized by a holomorphic superfield $\La$.
These models will generally generate a dynamical superpotential.
We expect a dynamical superpotential to be generated if there
is a holomorphic candidate that is invariant under all symmetries,
in particular the conformal $U(1)_R$.
Note that we are assuming that the Higgs couplings give mass
to all matter fields in the strong sector.
This means that there is a holomorphic mass scale $\La(H)$ for the
theory that depends on the Higgs fields.
For holomorphic quantities, $U(1)_R$ invariance is equivalent
to dimensional analysis, so there is always an allowed dynamical
superpotential of the form
\beq[Wdyn]
\De W_{\rm dyn} \sim \frac{\La^3(H)}{16\pi^2}.
\eeq
The most general form of the dynamical scale compatible with all
symmetries is
\beq[LambdaH]
\La^{6 - 2d}(H) \sim 16\pi^2 \ka_u^{2-d} \ka_d^{2-d} H_u H_d.
\eeq
In the explicit models of \S\ref{sec:modelscustodial} we can verify
that the usual dynamical superpotential is indeed generated and has
this form.
The factors of $4\pi$ in \Eqs{Wdyn} and \eq{LambdaH} follow from NDA.
We can understand them by noting that the interactions \Eq{CFTHiggsint}
are strong at the scale $\La$ in the limit
$\ep_{u,d} \to 1$, $4\pi v \to \La$.

If we replace the Higgs fields by their VEVs, we get a simpler and
more intuitive expression in terms of $\ep_{u,d}$:
\beq[epsilon]
\La \sim 4\pi (\ep_u \ep_d)^{1/2} (v_u v_d)^{1/2}
\sim 4\pi \ep v,
\eeq
which is consistent with Eq.~(\ref{eq:epsdef}). 
However, \Eq{LambdaH} must be used to get the correct form
of the $H_{u,d}$ dependence.

The superpotential \Eq{Wdyn} is non-analytic for at $H_{u,d} = 0$,
corresponding to the fact that it is obtained by integrating out
particles that get a mass from the Higgs VEVs.
An interesting special case is $d = \frac 32$, for which we have
\beq
W_{\rm dyn}(d = \sfrac 32)
\sim \ka_u^{1/2} \ka_d^{1/2} \, H_u H_d,
\eeq
\ie\ the superpotential is a pure $\mu$-term.
Note that this is what happens in the first model of
\S\ref{sec:modelscustodial} above.
For general $d$ the superpotential \Eq{Wdyn}
generates a Higgsino mass of order
\beq[Higgsinomassbootstrap]
\mu_{\rm dyn} \sim \frac{\d^2 W}{\d H^2} \sim \frac{\La^3}{(4\pi v)^2}
\sim \ep^2 \La.
\eeq
We will see below that this gives a
Higgsino mass term of order the Higgs mass,
giving a viable solution to the $\mu$-problem.
This is certainly an attractive feature of this class of models.

The supersymmetric contribution to the
Higgs potential from the dynamical superpotential has the form
\beq
V_{\rm dyn} \sim \frac{|\La(H)|^6}{(4\pi)^4}\,
\frac{H_u^\dagger H_u + H_d^\dagger H_d}{|H_u H_d|^2}
\sim H^{2d/(3-d)}.
\eeq
The potential is positive-definite, and
for $d > \frac 32$ the potential grows faster than $H^2$
so we can obtain a stable
electroweak breaking minimum by adding negative Higgs quadratic
terms.
We will be considering several different contributions to the Higgs 
potential that can balance each other in various combinations.
A good way to understand the relative sizes of the various
contributions is to look at the second derivative of the potential.
All of the contributions to the potential we will study go as a 
power of the Higgs fields.
The minimization is in general dominated by balancing two different
power-law contributions to the potential, so the second derivative
of each of them at the minimum will be the same up to factors of
order 1.
This is also equal to the physical Higgs mass (again up to factors
of order 1), so this allows us to estimate the physical
parameters associated with a given contribution to the potential,
assuming it is important for determining the Higgs VEVs.
In the present case, we have
\beq
V_{\rm dyn}'' \sim \frac{1}{(4\pi)^4}
(4\pi\ka^{2-d})^{6/(3-d)} v^{(4d - 6)/(3-d)}
\sim \frac{\La^6}{(4\pi v)^4}
\sim \ep^4 \La^2,
\eeq
where we have used the relations \Eqs{LambdaH} and \eq{epsilon}
in the last steps.
If this contribution to the
potential is important for stabilizing the Higgs mass,
we have
\beq
\ep 
\sim \left( \frac{m_h}{4\pi v} \right)^{1/3}.
\eeq
For numerical estimates, we use $4\pi v \simeq 2\TeV$, which is the value of the 
technirho mass in scaled-up QCD.
For $m_h \simeq 125\GeV$ we then obtain $\ep \sim 0.4$
and $\La \sim 800\GeV$.
In any stabilization where this contribution is important,
the Higgsino mass \Eq{Higgsinomassbootstrap} is
related to the Higgs mass by
\beq
\frac{\mu_{\rm dyn}}{m_h} \sim 1.
\eeq
In other words, the dynamical superpotential
gives a perfect parametric solution to the
$\mu$-problem.

The interactions in \Eq{CFTHiggsint} are rather close
to becoming strong at the scale $\La$.
This coincidence problem will be addressed in \S\ref{sec:coincidence}.
One may also worry about performing an expansion for small $\ep_{u,d}$
for such large values.
However, observables are suppressed by large powers of this
suppression, making the estimates more plausible.
However, these estimates clearly have large uncertainties,
and should be regarded as enlightened order-of-magnitude
estimates.

We now consider the contributions to the Higgs potential
from SUSY breaking in the strong sector.
There can be $A$ terms associated with the interaction
\Eq{CFTHiggsint} of the form
\beq[CFTAterms]
\De \scr{L} = \ka_u^{2 - d} A_u H_u \scr{O}_d 
+  \ka_d^{2 - d} A_d H_d \scr{O}_u
+ \hc
\eeq
These $A$ terms are not strongly renormalized by the strong sector,
since they are proportional to a relevant
coupling that is small in the UV.
It is therefore natural to have $A_{u,d} \sim \mbox{TeV}$.
The $A$ terms give a contribution to the Higgs potential
\beq[VAtermbootstrap]
V_A \sim \frac{\La^3(H)}{16\pi^2} (A_u + A_d) + \hc
\sim H^{3/(3 - d)}.
\eeq
Potential terms that are real parts of holomorphic functions
such as this always have unstable directions.
For $d < \frac 32$, \Eq{VAtermbootstrap} grows slower than $H^2$
and we can get a stable electroweak breaking minimum  by
balancing this against \emph{positive} SUSY-breaking $H^2$ terms
in the potential.
We have
\beq
V_A'' \sim \frac{A}{(4\pi)^2} ( 4\pi \ka^{2-d})^{3/(3-d)}
v^{(2d - 3)/(3 - d)}
\sim \frac{A \La^3}{(4\pi v)^2}
\sim \ep^2 A \La.
\eeq
If this contribution to the
potential is important for stabilizing the Higgs mass,
we have
\beq
\ep \sim \left( \frac{\La}{A} \right)^{1/4}
\left( \frac{m_h}{4\pi v} \right)^{1/2}.
\eeq
Note that $A \lsim \La$, otherwise
conformal symmetry breaking in the strong sector is dominated by
the $A$ terms.
The Higgsino mass \Eq{Higgsinomassbootstrap} is now
related to the Higgs mass as
\beq
\frac{\mu_{\rm dyn}}{m_h} \sim \ep \left( \frac{\La}{A} \right)^{1/2}.
\eeq
The Higgsino mass is not parametrically of order the Higgs
mass, but the values of $\ep$ are not very large, and we can
easily get a viable model without an additional contribution
to the Higgsino mass.

Other types of SUSY breaking terms in the strong sector
generally have large anomalous dimensions.
An exception are scalar mass-squared terms proportional to 
flavor generators (not including $U(1)_R$),
which are not renormalized.
If we assume that SUSY breaking originates at scales far above
the TeV scale, these are the only SUSY breaking terms in the strong
sector that are naturally of order the TeV scale.
(Other soft mass terms are generally suppressed at an
IR attractive fixed point, so this is a natural scenario.)

The non-renormalization of scalar mass-squared terms
proportional to flavor generators can be understood from the fact that
these mass terms can be written as the $D$-term for a
flavor gauge superfield.
Under conformal transformations, the flavor gauge fields
have dimension 0, so the soft mass-squared term have
dimension 2.
Similarly, the combination $\ka^{2-d} H$ is a chiral
primary field of dimension $3-d$.
This constraints how these fields can appear in the effective
theory below the scale $\La$.
(This is discussed in detail in the Appendix.)
The resulting potential can be expanded in powers of the soft
masses if these are a subleading contribution to conformal
symmetry breaking in the strong sector, and we get
\beq[Vsoftbootstrap]
V_{\rm soft} \sim \frac{1}{16\pi^2}
m_{\rm soft}^2 (4\pi \ka^{2-d} H)^{2/(3 - d)}.
\eeq
This contribution to the potential is not directly expressible
in terms of the holomorphic scale $\La(H)$, and the functional
form is not calculable (see Appendix).

In the models that we construct, 
the symmetries do not prevent a nontrivial potential
of this form.
We can choose this potential to stabilize or destabilize
$H = 0$ by choosing the sign of $m_{\rm soft}^2$.
Because we are assuming that the soft mass-squared terms
are a subleading contribution to conformal symmetry breaking
in the strong sector, negative mass-squared terms
will not induce a vacuum instability in the strong sector.
\Eq{Vsoftbootstrap} grows more slowly than $H^2$ for $d < 2$, so 
we can obtain a stable electroweak breaking minimum
by $m_{\rm soft}^2 < 0$ so that this contribution
destabilizes $H = 0$
and balancing it against a positive quadratic term for $H$.
The Higgs mass scale is then
\beq
V_{\rm soft}'' \sim m_{\rm soft}^2 \frac{\La^2}{(4\pi v)^2}
\sim \ep^2 m_{\rm soft}^2.
\eeq

To summarize, the contributions to the Higgs potential from
the strong sector are given by $V_{\rm dyn}, V_A, V_{\rm soft}$.
In the above we briefly discussed balancing each of these with 
tree-level Higgs mass terms
\beq
V_{\rm tree} = m_{Hu}^2 |H_u|^2 + m_{Hd}^2 |H_d|^2
+ B\mu (H_u H_d + \hc)
\sim m_H^2 H^2.
\eeq
However, other combinations are possible, and we will
summarize all possibilities.
We neglect the Higgs quartic terms from
the standard model gauge interactions, since these 
are far too small to give $m_h \simeq 125\GeV$.
\begin{itemize}
\item 
$V_{\rm dyn}$ and $V_{\rm tree}\,$:
This can work for $\frac 32 < d < 2$.
We obtain
\beq
\ep \sim \left( \frac{m_h}{4\pi v} \right)^{1/3} \sim 0.4,
\eeq
which gives $\La \sim 800\GeV$ for
$m_h \simeq 125\GeV$.
The dynamically generated Higgsino mass is related to the
Higgs mass by
\beq
\frac{\mu_{\rm dyn}}{m_h} \sim 1,
\eeq
so this gives a perfect parametric solution to the $\mu$-problem.
\item
$V_A$ and $V_{\rm tree}\,$:
This can work for $1 < d < \frac 32$.
We obtain
\beq
\ep \sim \left( \frac{m_h}{4\pi v} \right)^{1/2} 
\left( \frac{A}{\La} \right)^{-1/4}
\sim 0.25 \left( \frac{A}{\La} \right)^{-1/4},
\eeq
which gives $\La \sim 500\GeV \times (A/\La)^{-1/4}$.
We must have $A/\La \lsim 1$, since $A \sim \La$ 
corresponds to conformal symmetry breaking dominated by $A$.
Consistency requires that $V_{\rm dyn}$ is subdominant, which
occurs for
\beq[Alimit]
\frac{A}{\La} \gsim \left( \frac{m_h}{4\pi v} \right)^{2/3}
\sim 0.16.
\eeq
The dynamically generated $\mu$-term is given by
\beq
\frac{\mu_{\rm dyn}}{m_h} 
\sim \left( \frac{m_h}{4\pi v} \right)^{1/2}
\left( \frac{A}{\La} \right)^{-3/4}
\lsim 1
\eeq
where the bound follows from \Eq{Alimit}.
The dynamically generated
$\mu$-term is therefore parametrically too small in this
limit.
However, the suppression factors are not large, and it is
possible that the numerical value is sufficiently large.
\item
$V_{\rm soft}$ and $V_{\rm tree}\,$:
This can work for any $1 < d < 2$.
We obtain
\beq
\ep \sim \left( \frac{m_h}{4\pi v} \right)^{1/2}
\left( \frac{m_{\rm soft}}{\La} \right)^{-1/2}
\sim 0.25 \left( \frac{m_{\rm soft}}{\La} \right)^{-1/2}.
\eeq
We must have $m_{\rm soft} / \La \lsim 1$,
since $m_{\rm soft} \sim \La$
corresponds to conformal symmetry breaking dominated by $m_{\rm soft}$.
Consistency requires that $V_{\rm dyn}$ is subdominant, which
occurs for
\beq[softlimit]
\frac{m_{\rm soft}}{\La} \gsim \left( \frac{m_h}{4\pi v} \right)^{1/3}
\sim 0.4.
\eeq
The dynamically generated $\mu$-term is given by
\beq
\frac{\mu_{\rm dyn}}{m_h} \sim 
\left(
m_h \over 4 \pi v
\right)^{1/2}
\left( \frac{m_{\rm soft}}{\La} \right)^{-3/2}
\lsim 1,
\eeq
where the bound follows from \Eq{softlimit}.
\item
$V_{\rm dyn}$ and $V_A\,$:
This can work for $\frac 32 < d < 2$.
We obtain
\beq
\ep \sim \left( \frac{A}{\La} \right)^{1/2}
\sim \left( \frac{m_h}{4\pi v} \right)^{1/3}
\sim 0.4.
\eeq
Because $V_{\rm dyn}$ is part of the stabilization
of the potential, we have $\mu_{\rm dyn} \sim m_h$.
\item
$V_{\rm dyn}$ and $V_{\rm soft}\,$:
This can work for any $1 < d < 2$.
We choose $m_{\rm soft}^2 < 0$ to destabilize $H = 0$,
and $V_{\rm dyn}$ provides the stabilizing potential.
We obtain
\beq
\ep \sim \frac{m_{\rm soft}}{\La}
\sim \left( \frac{m_h}{4\pi v} \right)^{1/3}
\sim 0.4.
\eeq
We also have $\mu_{\rm dyn} \sim m_h$.
\item
$V_A$ and $V_{\rm soft}\,$:
This combination is not expected to work because
$V_A$ is always destabilizing and grows as a larger power
of $H$ than $V_{\rm soft}$.
\end{itemize}
The spectrum of the strong sector is approximately supersymmetric
in all of these cases, even the ones in which the Higgs VEVs are
determined by SUSY breaking interactions.
The reason is that we are always choosing parameters so that the
Higgs VEVs are the dominant source of conformal breaking in the strong
sector.
This is motivated by the fact that supersymmetric masses
in the strong sector naturally give a stable vacuum for the strong
sector.
If SUSY breaking dominates the breaking of conformal invariance
in the strong sector, this stability is lost in 
the simplest models
(recall the discussion in \S\ref{sec:modelscustodial}).

\subsection{Electroweak Preserving Masses
\label{sec:singletmass}}
We now consider the case where there is an electroweak-preserving
contribution to the mass scale in the strong sector that is the
dominant source of conformal symmetry breaking.
This can arise naturally from the VEV of a singlet Higgs field
$S$ coupled to the CFT via a superpotential interaction similar
to \Eq{CFTHiggsint}:
\beq
\De W = \ka_S^{2-d} S \scr{O}_S,
\eeq
where $\scr{O}_S$ is a CFT operator with the same dimension $d$
as $\scr{O}_{u,d}$.
Soft SUSY breaking naturally generates a VEV for $S$ that can
be somewhat larger than $v$.
It is also possible to have a $\mu$-like term $\De W \sim \scr{O}_S$,
but in general this would not be expected to give mass terms with
the same order of magnitude
as the contribution from the Higgs VEVs from \Eq{CFTHiggsint}.

In the limit where we neglect $\avg{H_{u,d}}$ the strong scale
is given by
\beq
\La_0^{3-d} \sim 4 \pi \ka_S^{2-d} \avg{S}.
\eeq
This is an arbitrary parameter of the theory.

The dominant source of conformal symmetry breaking in the strong
sector is assumed to come from the singlet VEV, which is
supersymmetric.
We therefore begin by analyzing the strong sector in the SUSY
limit.
The holomorphic strong scale is 
\beq[LambdaHsinglet]
\La(H) = \La_0 \left[ 1 + c_1 \ep_u \ep_d \frac{16\pi^2 H_u H_d}{\La_0^2}
+ c_2 \left( \ep_u \ep_d \frac{16\pi^2 H_u H_d}{\La_0^2} \right)^2
+ \cdots \right].
\eeq
where $c_{1,2}$ are order-1 numbers calculable in specific models.
The condition that the Higgs contribution to the dynamical scale
is subleading is then
\beq[deltaH]
\de_H \sim \left( \ep\, \frac{4\pi v}{\La_0} \right)^2
\lsim 1.
\eeq
This is the (square of) the
expansion parameter in \Eq{LambdaHsinglet}.
Note that having $\de_H < 1$ always requires some parametric
tuning, since minimizing the potential with
$c_{1,2} \sim 1$ gives $\de_H \sim 1$.
To get $\de_H \ll 1$ we need a smaller quadratic term, which
requires unnatural cancellations.
In fact, $\de_H$ is precisely the measure of fine tuning in
these models.
Just as in composite Higgs models, we are accepting a mild tuning
as the price for a model that has a light Higgs particle
and good precision electroweak fit.

We again have a dynamical superpotential of the form
\Eq{Wdyn}.
The corresponding supersymmetric
contribution to the Higgs potential is
\beq
V_{\rm dyn} &= \left|
\frac{1}{16\pi^2} \frac{\d \La^3(H)}{\d H} \right|^2
\nonumber\\
&\sim (\ep_u \ep_d)^2
\Bigl[ \La_0^2 (H_u^\dagger H_u + H_d^\dagger H_d)
\nonumber\\
&\qquad\qquad\qquad
{}+ 16\pi^2
\ep_u \ep_d (H_u^\dagger H_u + H_d^\dagger H_d) (H_u H_d + \hc)
+ \cdots
\Bigr].
\eql{Vdynsinglet}
\eeq
Note that we know the functional form of the potential
because we know the functional form of the holomorphic
scale $\La(H)$.
There are corrections to the $H$ \Kahler potential,
but they are smaller than the tree-level $H$ kinetic term.

We would like to consider the possibility that \Eq{Vdynsinglet}
gives the leading contribution to the Higgs quartic.
This scenario always requires some tuning, because
it requires the $H^2$ terms to be somewhat smaller
than in \Eq{Vdynsinglet} so that the Higgs VEV is a
subleading contribution to $V_{\rm dyn}$.
However, the model can work with only very mild tuning.
If this potential is part of the stabilization of the Higgs
VEVs, the Higgs mass parameter is
\beq[mHsinglet]
V_{\rm dyn}'' \sim 
(\ep_u \ep_d)^{3} 
\left(
4\pi v
\right)^2 .
\eeq
For a Higgs mass of $125\GeV$ this gives 
$(\ep_u \ep_d)^{1/2} \sim 0.4$ as above.
The measure of tuning is precisely the expansion parameter
$\de_H$ in \Eq{deltaH}.
There is a trade-off between naturalness and predictability,
but we get a plausible scenario for moderately small values
of $\de_H$, \eg\ $\de_H \sim 0.2$.

The quartic Higgs interaction in \Eq{Vdynsinglet}
is not positive definite.
The full dynamically generated potential is positive definite, 
so the question is  whether there is a stable electroweak-breaking
minimum with VEVs sufficiently small
that the quartic term dominates.
It is easy to see there is such a local minimum
even in the limit where the SUSY
quartic vanishes, for a sufficiently large
$B\mu$ term of the correct sign.
Specifically, for the potential
\beq\bal
V &= m_{Hu}^2 H_u^\dagger H_u + m_{Hd}^2 H_d^\dagger H_d
- B\mu (H_u H_d + \hc)
\\
&\qquad
{}+ \la (H_u^\dagger H_u + H_d^\dagger H_d) (H_u H_d + \hc)
\eal\eeq
we find a local stable electroweak breaking minimum for $H_u H_d > 0$
provided that
\beq
B \mu > \frac{5\la v^2}{8} \left[ 1 + \frac 15 \cos(4\be)\right].
\eeq
This solution is in general only a local minimum, since there is
an unstable $D$-flat direction $|H_u| = |H_d|$, $H_u H_d < 0$.
The global minimum can be far away in field space, so at least
some of these solutions are expected to be cosmologically acceptable.
We will not explore this issue further here.

The dynamical superpotential also generates a Higgsino mass
\beq
\mu_{\rm dyn} \sim \ep_u \ep_d \La_0.
\eeq
\Eq{mHsinglet} then implies that
\beq
\frac{\mu_{\rm dyn}}{m_h} \sim \frac{1}{\de_H^{1/2}}.
\eeq
The dynamically generated
$\mu$-term is somewhat larger than the Higgs mass,
which still gives a good solution to the $\mu$-problem.

We now turn to SUSY breaking contributions to the potential.
Because we are assuming that the Higgs VEVs are a subleading contribution
to the potential, the potential can always be expanded in terms
of gauge-invariant combinations of Higgs fields, so the
dominant contributions will be qualitatively similar to
$V_{\rm dyn}$ discussed above.
In particular, we always require a tuning of order $\de_H$
given by \Eq{deltaH} to make the Higgs contribution subleading.

Specifically, we concerned with $A$ terms of the form
\beq
\De \scr{L} = \ka_S^{2 - d} A_S S \scr{O}_S + \hc
\eeq
and scalar mass-squared terms $m_{\rm soft}^2$
proportional to flavor generators.
These give rise to a potential of the form
\beq[Vsoftsinglet]
V_{A,{\rm soft}} \sim \xi_{A,{\rm soft}}
\frac{\La_0^4}{16\pi^2} \left[
1 + \left( \frac{4\pi \ep H}{\La_0} \right)^2
+ \left( \frac{4\pi \ep H}{\La_0} \right)^4
+ \cdots
\right].
\eeq
We assume that the SUSY breaking terms are a subleading contribution
to conformal symmetry breaking in the strong sector, so we can
expand the suppression factor in powers of the SUSY breaking masses:
\beq
\xi_A &\sim \frac{A_S}{\La_0},
\\
\eql{xisoft}
\xi_{\rm soft} &\sim \frac{m_{\rm soft}^2}{\La^2_0}.
\eeq
$V_A$ is the real part of a holomorphic function, and therefore
cannot stabilize the Higgs VEV.
(The $H^2$ terms in $V_A$ are proportional to
$H_u H_d + \hc$ and the quartic terms are proportional
to $(H_u H_d)^2 + \hc$)
As with $V_{\rm dyn}$, we need parametric tuning of order $\de_H$
given by \Eq{deltaH} to make the Higgs VEV contribution subleading.

We will assume that either $V_{\rm dyn}$ or $V_{\rm soft}$
dominates the Higgs quartic.
We then have
\beq
m_h^2 \sim \la v^2 \sim 16\pi^2 \xi_{\rm dyn,soft} \ep^4 v^2
\eeq
where $\xi_{\rm soft}$ is given in \Eq{xisoft} and
\beq
\xi_{\rm dyn} \sim \ep^2.
\eeq
The two possible origins for the Higgs quartic are
then as follows.
\begin{itemize}
\item
$V_{\rm dyn}\,$:
This requires
\beq
\ep \sim \left( \frac{m_h}{4\pi v} \right)^{1/3}
\sim 0.4.
\eeq
The scale $\La_0$ depends on the degree of fine-tuning:
\beq
\La_0 \sim \de_H^{-1/2} \ep 4\pi v \sim 800\GeV \times \de_H^{-1/2}.
\eeq
\item
$V_{\rm soft}\,$:
This requires
\beq
\ep \sim \left( \frac{m_h}{4\pi v} \right)^{1/2} 
\left( \frac{m_{\rm soft}}{\La_0} \right)^{-1/2}
\sim 0.25 \left( \frac{m_{\rm soft}}{\La_0} \right)^{-1/2}.
\eeq
Requiring that the soft mass contribution dominates $V_{\rm dyn}$,
we obtain
\beq
\frac{m_{\rm soft}}{\La_0} \gsim 
\left( \frac{m_h}{4\pi v} \right)^{1/3} \sim 0.4.
\eeq
We then have
\beq
\La_0 \sim 500\GeV \times \de_H^{-1/2} 
\left( \frac{m_{\rm soft}}{\La_0} \right)^{-1/2}
\lsim 800\GeV \times \de_H^{-1/2}.
\eeq
\end{itemize}

\subsection{Coincidence Problem
\label{sec:coincidence}}
A potential problem with this framework is that 
$\ka_{u,d}$ in \Eq{CFTHiggsint} are dimensionful
parameters that must be near the TeV scale in order to have a
successful model.
This is analogous to the $\mu$-problem in the MSSM,
where a SUSY-preserving mass must be of order the SUSY breaking
masses.
We have seen that in our models a Higgsino mass of the correct
size can be dynamically generated, so there is no need for a $\mu$-term,
so we have traded one coincidence problem for another.

Here we point out that the mass scale $\ka_{u,d}$ can naturally
be near the TeV scale by a
generalization of the Giudice-Masiero mechanism \cite{SCTC}.  
We focus on the minimal model
discussed in \S\ref{sec:modelscustodial} and let us
consider a SUSY breaking field $X$ with a nonzero $F$-term $\langle F_X
\rangle$ in the hidden sector.  The visible sector SUSY breaking
including $m_{\rm soft}$ and $m_H$ is given by
\begin{equation}
M_{\rm SUSY} \sim \frac{\langle F_X \rangle}{M_{\rm med}},
\end{equation}
where $M_{\rm med}$ is the mediation scale.
In addition, we assume that the hidden sector contains a field $Y$ with the following expectation values:
\begin{equation}
\langle Y \rangle \sim \langle F_X \rangle^{1/2},
\end{equation}
and $\avg{F_Y}$ sufficiently small.
Then the interaction \Eq{CFTHiggsint} can be generated by
the superpotential
\begin{equation}
W_{\rm eff} \sim \frac{1}{M_{\rm med}^{1/2}} Y H \Psi_i \tilde{\Psi}_j,
\end{equation}
and we have $\kappa \sim M_{\rm SUSY}$.  
The correct size of $\langle Y \rangle$
can be realized with the hidden sector superpotential:
\begin{equation}
W_{\rm hid} = c_1 X + \frac{c_2}{M_{\rm med}} Y^4,
\end{equation}
where $c_{1,2}$ are order-1 coupling constants.

Another possibility to solve the coincidence problem
is to consider strong sectors where the operators $\scr{O}_{u,d}$
have dimension $d \simeq 2$ \cite{Vafa}.
This does not occur in strongly-coupled
SUSY QCD models, but is perfectly allowed in more general
strong theories.
In such models, the couplings $\ka_{u,d}$ are nearly
dimensionless, requiring a milder coincidence.
However, we have seen that we need this coupling to be
rather strong at the strong scale $\La$:
$\ep \sim 0.4$, meaning that the coupling is about half
its strong-coupling value.
This means that in general the coupling is running fast at
the scale $\La$, and we still have a coincidence problem.
Alternatively, we could assume that the couplings $\ka_{u,d}$
are slightly relevant and approach a strong fixed point,
giving a robust explanation for the large value of the coupling.
At this fixed point the Higgs fields $H_{u,d}$ become conformal
operators with dimension near 1.
This implies that they are nearly
decoupled from the CFT, in contradiction with what is required
for phenomenology.
We conclude that it is not clear whether $d \simeq 2$ really
gives a viable theory without a coincidence problem.

\section{Precision Electroweak Tests
\label{sec:PEW}}
We now consider the constraints on this scenario from
precision electroweak tests.
Our strong sector does not couple directly to the quarks and
leptons, and therefore the only important corrections are the
$S$ and $T$ parameters
\cite{Peskin}.
(In particular, the correction to $Z \to \bar{b}b$ is negligible.)
These are simply effective operators that arise
from integrating out the strong sector.
We cannot perform a precise computation, but we can estimate them using
NDA.

\subsection{$T$ Parameter}
We now estimate the $T$ parameter, which gives the strongest
constraint on this class of models.

We first consider models without electroweak-preserving masses
in the strong sector, as discussed in \S\ref{sec:Higgsbootstrap}.
We assume that the strong sector has a custodial symmetry in
the limit where the Higgs couplings are turned off,
as in the models of \S\ref{sec:modelscustodial}.
In general, the Higgs couplings do not preserve the custodial
symmetry ($\ka_u \ne \ka_d$), and it is not natural to have
$v_u = v_d$.
This means that custodial symmetry is generically
broken in the strong sector.
This leads to large corrections to the $T$ parameter, but these
can be made sufficiently small with mild tuning.
The strong sector will have custodial symmetry if 
accidentally $\ka_u^{2-d} v_u \simeq \ka_d^{2-d} v_d$.
The tuning required is measured by
\beq
\de_C = \frac 12 \,
\frac{\ka_u^{2-d} v_u - \ka_d^{2-d} v_d}
{\ka_u^{2-d} v_u + \ka_d^{2-d} v_d}.
\eeq
The contribution to the $T$ parameter from the strong sector
is estimated by the contribution of $N$ doublets of fermions
with electroweak breaking masses of order $\La$,
with custodial breaking mass splitting of order 
$\De m \sim \de_C \La$.
This is similar in spirit to estimating the short-distance contribution to
the $S$ parameter in technicolor models using the contribution from 
``constituent techniquarks,''
which is known to give an accurate answer both parametrically and
numerically.
In the present case, this gives
\beq
\De T \simeq \frac{N (\De m)^2}{12\pi s_W^2 c_W^2 m_Z^2}
\sim 0.4 \left( \frac{N}{4} \right) 
\left( \frac{\ep}{0.4} \right)^2
\left( \frac{\de_C}{0.1} \right)^2.
\eeq
We see that the corrections to the $T$ parameter are moderate 
at the price of roughly $10\%$ tuning.

We next consider the case where the electroweak-preserving
masses in the strong sector dominate.
The important parameter in this case is $\de_H$,
the square of the ratio of the electroweak-preserving and
electroweak breaking masses (see \Eq{deltaH}).
This ratio can only be made small by tuning, and $\de_H$ is a
measure of this tuning.
In this case, we estimate the tuning by $N$ doublets of fermions
with electroweak breaking mass $\De m \sim \de_H^{1/2} \La_0$
\cite{Lavoura}:
\beq
\De T \simeq \frac{13}{480\pi s_W^2 c_W^2 m_Z^2}
\frac{N \De m^4}{\La_0^2}
\sim 1.2 \left( \frac{N}{4} \right)
\left( \frac{\ep}{0.4} \right)^2
\left( \frac{\de_H}{0.1} \right),
\eeq
where we have used
\beq
\frac{(\De m)^4}{\La_0^2} \sim \de_H^2 \La_0^2 
\sim \de_H \ep^2 (4\pi v)^2.
\eeq
Note that the reduction in $\De T$ is only linear in $\de_H$,
so it parametrically more tuning is required to get a sufficiently
small $T$ parameter in this case.
Of course, all of these estimates have order-1 uncertainties,
but the tuning required to get
a phenomenologically viable $T$ parameter is significantly less than
the $1\%$ (or less) tuning of the MSSM.

\subsection{$S$ Parameter}
We estimate the contribution to the $S$ parameter from $N$
doublets of fermions with electroweak breaking masses.
We then obtain
\beq
\De S \sim \frac{N}{6\pi} \sim 0.2 \left( \frac{N}{4} \right).
\eeq
In the case where electroweak preserving masses dominate, the $S$
parameter is suppressed by a factor of $\delta_H$ compared to the above
estimate.

We expect the strong sector
contributions to both the $S$ and $T$ parameters
to be positive.
This is true for perturbative contribution from electroweak breaking
multiplets, and also holds for holographic theories where
$S$ and $T$ are calculable.
This direction allows the largest deviation from the standard
model in the $S$-$T$ plane.
It is also worth keeping in mind that it is quite possible that
the estimates for the $S$ and $T$ parameters above are too large,
either because our estimate for $\ep$ is too large
or because the order-1 factors are small.
We conclude that it is very plausible
that a good fit can be obtained in these models
at cost of a moderate tuning.

\section{LHC signatures}

In this subsection we consider the signatures of this model at the LHC.

\subsection{$h \to \ga\ga$ decay
\label{sec:hgg}}
The strong sector contains charged particles with a substantial
coupling to the Higgs boson, and therefore gives an important
contribution to $h \to \ga\ga$.
This is one of the important channels in which the $125\GeV$ Higgs
signal is present, and will be probed in the coming year of LHC
running.
Because the strong sector is approximately supersymmetric,
the leading correction to $h \to \ga\ga$ is in fact calculable
\cite{Heckman}.
This is because in the low-energy effective theory below
the scale $\La$, the strong sector contribution can be parameterized
by the effective interaction
\beq
\scr{L}_{\rm eff} = -\frac{1}{4 e^2} F^{\mu\nu} F_{\mu\nu},
\eeq
The low-energy
effective coupling depends logarithmically on $\La(H)$ due to
the renormalization group.
Above the scale $\La(H)$, the renormalization group equation is
given by the NSVZ beta function
\beq
\mu \frac{d}{d\mu} \left( \frac{1}{e^2} \right)
= -\frac{1}{16\pi^2} \sum_r Q_r^2 (1 + \ga_r),
\eeq
where
\beq
\ga_r = \mu \frac{d}{d\mu} \ln Z_r
\eeq
is the anomalous dimension of the chiral superfield $r$.
The anomalous dimension term is negligible except for the strongly
coupled fields.
Assuming that the operators $\scr{O}_{u,d}$ in \Eq{CFTHiggsint} are bilinear
``meson'' operators of a SUSY QCD theory,
the anomalous dimension is related to the
dimension of the operators by
\beq
d = 2 - \ga.
\eeq
The beta function is therefore a constant above and below
the threshold $\La$, and the low-energy coupling depends on
the scale $\La(H)$ via the difference in the beta functions
above and below the scale $\La(H)$:
\beq
\frac{1}{e^2} = -\frac{3-d}{16\pi^2} \left( \sum_r Q_r^2 \right) 
\ln \La(H) + \cdots.
\eeq
For example, in the minimal model of \S\ref{sec:modelscustodial}
we have
\beq
\sum_r Q_r^2 = 8(1 + 4b^2),
\qquad
d = \frac 32,
\eeq
where $b$ is the $B - L$ charge.

In models where the dynamical scale is determined entirely by the
Higgs VEVs, we have (see \Eq{LambdaH})
\beq
\La^{6 - 2d}(H) = \La_0^{6 - 2d} \frac{H^0_u H^0_d}{v_u v_d},
\eeq
where $\La_0$ is the value of the dynamical scale when
the Higgs fields are replaced by their VEVs.
For models where the masses in the strong sector are dominantly
electroweak singlet, we have
\beq
\La(H) = \La_0 \left[ 1 + \de_H \frac{H^0_u H^0_d}{v_u v_d} 
+ \cdots \right].
\eeq
For the first type of model we have
\beq
\scr{L}_{\rm eff} = \frac{e^2}{16\pi^2}
\, \frac{1}{4} \left( \sum_r Q_r^2 \right)
\frac{\cos(\al + \be)}{\sin 2\be}
\times \frac{h}{v} F^{\mu\nu} F_{\mu\nu},
\eeq
where we have canonically normalized the photon fields
and used the standard definitions
\beq
H_u^0 = \frac{1}{\sqrt{2}} \left( v \sin\be + h \cos\al + \cdots \right),
\\
H_d^0 = \frac{1}{\sqrt{2}} \left( v \cos\be - h \sin\al + \cdots \right).
\eeq
The decay width is then given by
(see \eg\ \Ref{Spira})
\beq
\Ga(h \to \ga\ga) \propto
\left| A_{\rm SM} + \sfrac 12 \left( \sum_r Q_r^2 \right)
\frac{\cos(\al+\be)}{\sin 2\be} \right|^2,
\eeq
where
$A_{\rm SM} \simeq -6.5$.
For the second type of model, we obtain the same effective coupling
multiplied by an additional factor of $2(3 - d) \de_H$.

For the minimal model with
$b=0$ and $\cos (\alpha + \beta) / \sin 2 \beta \sim 1$, the width
is reduced by $\frac 17$.
On the other hand, with $b=1$, there is
an enhancement by a factor of $4$.
The charge assignments in the strong sector are rather restricted
if we impose the requirement that there are no states with fractional
electromagnetic charges, which would give rise to cosmologically 
dangerous stable charged particles.
For example, having only one pair of electroweak doublets
in the $SU(2)$ model, the smallest $B - L$ charge we can have
for the electroweak singlets is $\pm 1/2$, and we obtain the
same value for $h \to \ga\ga$.
However, in models with singlets, the correction to
$h \to \ga\ga$ may be significantly reduced at the cost of
a mild tuning.
We conclude that generically we expect a significant deviation from the
standard model rate for $h \to \ga\ga$,
and either an increase or decrease is possible.

\subsection{TeV Resonances}
The strong sector has hadronic resonances at the scale
$\La \sim 800\GeV$.
This scale is significantly smaller than the scale
$4\pi v \sim 2\TeV$ we expect for a technicolor theory
because the Higgs is weakly coupled to the strong sector.
In this section, we outline some promising LHC
signals of these resonances.

The strong sector is approximately supersymmetric, so the
heavy resonances will appear in approximately degenerate
supermultiplets.
For example, the vector multiplet will also contain a 
scalar and a fermion.
If $R$-parity is a symmetry, there will be $R$-parity
even and odd resonances.
The $R$-parity even resonances can be singly produced,
and we will focus on their phenomenology here.

We focus on models where
the strong sector has an $SU(2)_L \times SU(2)_R$ global 
symmetry that is weakly gauged by the standard model.
To get a good precision electroweak fit, the masses
induced by the Higgs VEVs must be chosen to be
approximately invariant under the custodial $SU(2)_C$,
the diagonal subgroup of $SU(2)_L \times SU(2)_R$.
The full $SU(2)_L \times SU(2)_R$ is nonlinearly realized
by the Higgs fields $H_{u,d}$ at the strong scale $\La$.
In fact, since we are considering a supersymmetric theory,
we have a nonlinear realization of
the complexified group~\cite{Zumino:1979et}.  Couplings to matter fields
in arbitrary representations of the (complexified) $SU(2)_C$ are
discussed in \Ref{Bando:1983ab}.

In models with unification (but no custodial symmetry),
the strong fields come in complete $SU(5)$ multiplets.
A good precision electroweak fit requires that the masses
in the strong sector be approximately electroweak singlet,
but the doublet and triplet masses need not be approximately
the same.
The resonances in the strong sector therefore come in
complete multiplets of the standard model gauge group
$SU(3)_C \times SU(2)_W \times U(1)_Y$.

We will focus our discussion in this section
on the models with custodial symmetry,
since these require less tuning.
A natural candidate for the lightest resonance would be
a vector supermultiplet in the $\bf{1}+\bf{3}$ representation of
$SU(2)_C$, which would contain the singlet scalar $\sigma$, the triplet
scalar $a_0$ and the analog of the QCD $\omega$ and $\rho$ mesons.
We will not make the group theory structure explicit in the formulas
below, since we are interested only
in order-of-magnitude estimates.

Symmetries allow a presence of  
a kinetic mixing term between the Higgs fields and the
singlet scalar component $\si$ of the resonance multiplet:
\beq
\De\scr{L} \sim \frac{\La}{4\pi v} \d^\mu h \d_\mu \si
\sim \ep \d^\mu h \d_\mu \si.
\eeq
Electrically neutral scalar resonances can therefore be produced by
gluon fusion with a cross section
of order $\ep^2 \sim 0.1$ times the gluon fusion
cross section for producing
a standard-model Higgs of the same mass.
Through a similar mixing term for the charged Higgs boson, charged
scalar resonances $a_0^\pm$ can be produced by $g b \to a_0^- t$ with a
cross section of order $\ep^2$ times the cross section for a reference
charged elementary Higgs of the same mass \cite{TCscalarpheno}.
For vector resonances, there can be a mixing term
between standard model vector bosons $V_\mu$
($Z_\mu$, or $W_\mu$)
and the vector component $\rho_\mu$ of the resonance
of the form
\beq
\De\scr{L} \sim \frac{g \La^2}{4\pi} \, V^\mu \rho_\mu
\eeq
where $g$ is the $V_\mu$ coupling constant.
This allows single production of vector resonances
via Drell-Yan processes with a mixing suppression
of order $g/4\pi$.
The production rate is therefore $g^2 / (4\pi)^2 \sim 10^{-2}$
times that of a sequential $W'$ or $Z'$ of the same
mass.

We now discuss decays.
The largest couplings are to the Higgs fields, and longitudinal
$W$ and $Z$ particles which are equal to those of the corresponding
Goldstone fields by the equivalence theorem.
The matrix elements of the decays of either the scalar or vector
mesons to pairs of Higgs fields or Goldstones are all given by
\begin{equation}
\mathcal{M} \sim \frac{\Lambda^3}{4\pi v^2} 
\sim 4 \pi \epsilon^2 \Lambda.
\end{equation}
The width of the heavy resonances is therefore suppressed by
$\ep^4 \sim 10^{-2}$, and we expect them to be narrow.
If we can neglect phase space suppression, we expect the decays to
different Higgs particles and Goldstones to be comparable.
A particularly interesting final states include
$AA$ followed by $A \to Z h$ followed by either 
$h \to WW$ or $h \to \bar{b}b$.
But $W_L W_L$, $Z_L Z_L$, $hh$, and $H^+ H^-$ are all
worth further study.
The decay to photons is highly suppressed because
photons do not have longitudinal polarizations:
$\Ga_{\ga\ga} / m \sim (e / 4\pi)^4 \sim 10^{-4}$.

We note that this class of models is another case where
a stronger Higgs sector leads to
resonant production of heavy standard-model particles.
The 2-Higgs doublet model is a good simplified model
for many of these searches
\cite{simpstrong}.

The LHC experiments have already put constraints on such resonances
from resonant $WZ$ production followed by decays into 
3 leptons~\cite{Aad:2012vs, Collaboration:2012kk}.
The bound on the cross section times the branching ratio
obtained there can be translated into the constraints on the $\rho$-like
vector resonance.
This has been done in, for example, Refs.~\cite{Falkowski:2011ua,
Bellazzini:2012tv}. 
By comparing with the corresponding parameter points
($g_\rho \sim 4 \pi$ and $g_{\rho \pi \pi} \sim 4 \pi \epsilon^2$ in
Ref.~\cite{Falkowski:2011ua}, or $m_\rho \sim 800$~GeV and $a_\rho \sim
\epsilon$ in Ref.~\cite{Bellazzini:2012tv}), we can see that the point
is close to the current LHC bound. 
Therefore, the discovery of such a resonance could happen soon in these
models.

\section{Conclusions
\label{sec:conclusion}}
In this paper we have proposed models that address the naturalness
problem of supersymmetric models by partial Higgs compositeness.  The
models consist of the MSSM or NMSSM coupled to a strong conformal 
sector via standard model gauge interactions and Higgs couplings of 
the form \Eq{CFTHiggsint}.
The strong superconformal sector allows large Higgs couplings
without Landau poles in the UV.
Vacuum expectation values of the elementary Higgs fields $H_{u,d}$
and/or singlet Higgs fields break conformal symmetry in the strong
sector.
The strong sector has a mass scale of order $800\GeV$,
and generates corrections to the Higgs potential that can
explain a 125~GeV Higgs mass without any fine-tuning.
The Higgs coupling also generates a Higgsino mass of order the
Higgs mass, so there is no need for a $\mu$-term.
On the other hand, the Higgs couplings are dimensionful relevant
couplings, and the fact that they are near the TeV scale
requires an explanation.
As an example, we show that a generalization of the Giudice-Masiero
mechanism can give a natural explanation.

Unlike technicolor theories, the Higgs fields are only partially
composite so that we can naturally have a top-quark Yukawa coupling
that is order 1.
Even though electroweak symmetry is broken by the VEVs of weakly-coupled
Higgs fields, there are in general large corrections to the $T$ parameter
coming from the Higgs VEVs, which must be tuned away to get a good
precision electroweak fit.
The tuning is of order $10\%$, significantly better than the $1\%$
(or worse) tuning in typical supersymmetric models.

The models predict a rich phenomenology that differs in interesting
ways from that of the MSSM.
There are generally large corrections to the $h \to \ga\ga$
width, either suppression or enhancement depending on the charge
assignments of the strong sector.
There is also an approximately supersymmetric spectrum of hadrons
at the scale $\sim 800\GeV$ that decay to pairs of MSSM Higgs
particles or longitudinally polarized $W$ and $Z$s.
We believe that this is a plausible framework for natural
supersymmetry that is well worth further exploration.

\section*{Acknowledgments
\label{sec:Acknowledgments}}

We would like to thank A. Hebecker for discussions.  
MAL is supported by the US Department of Energy under grant
DE-FG02-91-ER40674.
RK is supported in part by
the Grant-in-Aid for Scientific Research 23740165 of JSPS.  
YN is supported by JSPS Fellowships for Young
Scientists.
YN would like to acknowledge a great debt to his former home institute,
the Yukawa Institute for Theoretical Physics, Kyoto University 
where most of his work was done.

\appendix{Appendix: SUSY Breaking in the Conformal Sector}
In this appendix, we discuss the contribution of soft SUSY breaking
terms in the strong sector to the Higgs potential.
Specifically, we consider mass-squared terms proportional to
generators of the flavor group and exhibit the conditions under
which these lead to a contribution to the Higgs potential of the
form \Eqs{Vsoftbootstrap} and \eq{Vsoftsinglet}.

We consider $SU(N)$ SUSY QCD in the conformal window,
\ie $\frac 32 N \le F < 3N$, where $F$ is the number of flavors.
We begin with the case $N \ge 3$, and discuss $N = 2$
separately below.
We add mass terms for all quarks:
\beq[Massterm]
W = \tilde{Q}^T M Q.
\eeq
In our models, $M$ is proportional to Higgs and/or singlet fields,
and the scale of conformal symmetry breaking 
will be determined self-consistently
by the VEVs of these fields.

A straightforward way to see that the symmetries of the problem do not
prevent a potential of the form 
\Eqs{Vsoftbootstrap} and \eq{Vsoftsinglet}
is to compute the 1-loop potential in the perturbative case.
To be concrete, we consider a model with custodial symmetry:
\begin{eqnarray}
 W = \sum_i y_i H \tilde Q^i Q^i.
\end{eqnarray}
The chiral superfields $Q$ and $\tilde Q$ are fundamental and
anti-fundamental representation of the $SU(N)$ gauge group, and $i$ runs
from one to $F/2$.
The one-loop effective potential is given by
\begin{eqnarray}
 V_{\rm 1-loop} = 
{N \over (4 \pi)^2} \sum_i 
&&\left[
{\rm tr} \left \{
{1 \over 2}
(y_i^2 H^\dagger H + m_i^2)^2
\left(
\ln {
y_i^2 H^\dagger H + m_i^2
 \over \mu^2 }
-{\frac 32}
\right)
\right \}
\right.
\nonumber \\
&&
+
\left.
{\rm tr} \left \{
{1 \over 2}
(y_i^2 H^\dagger H + \tilde m_i^2)^2
\left(
\ln {
y_i^2 H^\dagger H + \tilde m_i^2
 \over \mu^2 }
-{\frac 32}
\right)
\right \}
\right.
\nonumber \\
&&
- 
\left.
{\rm tr} \left \{
(y_i^2 H^\dagger H)^2
\left(
\ln {y_i^2 H^\dagger H \over \mu^2 }
-{\frac 32}
\right)
\right \}
\right].
\end{eqnarray}
Here we added soft masses $m_i^2$ and $\tilde m_i^2$ for $Q_i$ and
$\tilde Q_i$, respectively. The scale $\mu$ can take an arbitrary value.
The contribution at the linear order in $m_i^2$ or $\tilde m_i^2$ is given by
\begin{eqnarray}
 V_{\rm 1-loop} 
= 
\sum_i {\rm tr} \left\{
{y_i^2 N \over (4 \pi)^2}  (m_i^2 + \tilde m_i^2)
|H|^2
\left(
\ln 
{y_i^2 |H|^2 \over \mu^2} 
- 1
\right) \right\}
+ O(m^4).
\end{eqnarray}
One can see that the potential is generically non-vanishing even when
the soft masses are proportional to generators of the flavor
group. 
In particular, it survives when
\begin{eqnarray}
 \sum_i (m_i^2 + \tilde m_i^2) = 0,
\end{eqnarray}
for a generic set of $y_i$.

We can get a better understanding of this with a non-perturbative
analysis based on symmetries, as follows.
Below the scale $M$, all the quarks get massive, and the
low-energy theory is pure SUSY Yang-Mills.
This theory confines, generating a dynamical scale $\La$.
We have
\beq
d = \dim(\tilde{Q}Q) = 3 \left( 1 - \frac{N}{F} \right) 
\eeq
and therefore
\beq[Lambdappendix]
\La \propto (\det M)^{1/F(3 - d)}
= (\det M)^{1/3N}.
\eeq
In a strongly coupled theory ($F \simeq 2N$) the proportionality
constant will be order 1.

We now consider the effects of soft SUSY breaking in the strong
sector.
The only soft SUSY breaking terms that are not renormalized
are mass-squared terms proportional to generators of the
flavor group $SU(F) \times SU(F) \times U(1)$.
These can be parameterized by the $D$-term of
$SU(F) \times SU(F) \times U(1)$ gauge superfields
$X, \tilde{X}$
\beq
X = \th^4 m^2,
\qquad
\tilde{X} = \th^4 \tilde{m}^2.
\eeq
The flavor gauge fields must satisfy
\beq
\tr(X + \tilde{X}) = 0
\eeq
to project out the anomalous $U(1)$.
We write the kinetic term as
\beq
\scr{L}_{\rm kin} = \myint d^4\th \left[
Q^\dagger e^{X} Q + \tilde{Q}^\dagger e^{\tilde{X}} \tilde{Q}
\right],
\eeq
where we suppress the gauge fields.
The theory is therefore invariant under
the flavor gauge transformations
\beq
Q \mapsto U Q,
\qquad
\tilde{Q} \mapsto \tilde{U} \tilde{Q}
\eeq
and
\beq
e^{-X} \mapsto U e^{-X} U^\dagger,
\qquad
e^{-\tilde{X}} \mapsto \tilde{U} e^{-\tilde{X}} \tilde{U}^\dagger
\eeq
provided we let $M$ transform as
\beq
M \mapsto \tilde{U}^{T -1} M U^{-1}.
\eeq
Here $U, \tilde{U}$ are complexified flavor transformations,
so $U^\dagger \ne U^{-1}$.

We are interested in the effective \Kahler potential below
the scale $\La$.
This must be invariant under the flavor gauge transformations
above.
Note that
\beq[Sdefn]
S = e^{-X} M^\dagger e^{-\tilde{X}^T} M \mapsto U S U^{-1}.
\eeq
We can use this to construct the invariants
\beq[Sn]
\tr(S^n) = \tr[(M^\dagger M)^n]
- n \tr[ X(M^\dagger M)^n + \tilde{X}^T (M M^\dagger)^n]
+ O(X^2).
\eeq
The most general dynamical \Kahler potential is a homogeneous
function of these variables with the degree of homogeneity fixed
by dimensional analysis.
As long as the symmetries of the mass terms $m^2, \tilde{m}^2$
and the mass term $M$ do not force the $O(X)$ term in \Eq{Sn}
to vanish, we expect a nonzero
potential of the form \Eqs{Vsoftbootstrap} and \eq{Vsoftsinglet}.

We also need to take into account the constraints of superconformal
invariance.
The mass $M$ in \Eq{Massterm} is a chiral primary field with dimension
$3 - d$.
We can write a superconformal invariant kinetic term for a chiral
primary field $\Phi$ with dimension $1$ as 
$\int d^4\th\, \Phi^\dagger \Phi$. 
However, we cannot define $\Phi$ as a power of the matrix $M$ because this 
does not have a well-defined transformation under flavor
gauge symmetries.
The holomorphic scale $\La$ is a chiral superfield of dimension 1
(see \Eq{Lambdappendix}), but it is a singlet under the flavor symmetry, 
and $\int d^4\th\, \Lambda^\dagger \La$ does not depend on the flavor gauge
fields $X, \tilde{X}$.
However, we can define the quantity
\beq
\De^a{}_b = \frac{\d \det(M)}{\d M_{\tilde{a}a}} M_{\tilde{a} b},
\eeq
where $a, b, \ldots$ and $\tilde{a}, \tilde{b}, \ldots$ are
$U(F) \times U(F)$ flavor indices.
$\De \sim M^F$ is a chiral superfield that transforms under 
flavor in the adjoint representation, \ie
$\De \mapsto U \De U^{-1}$.
The field
\beq
\Phi = \De^{1/F(3 - d)} \sim \De^{1/3N}
\eeq
is therefore a chiral primary field of dimension 1,
and we can write the conformally invariant and
flavor gauge invariant kinetic term
\beq
K_{\rm eff} = \tr(\Phi^\dagger e^X \Phi e^{-X}).
\eeq
Note that in 4D the kinetic term for a complex scalar with
gauge interactions is conformally invariant, with the gauge
fields dimensionless.
Note that $K_{\rm eff}$ can be written in terms of the invariants
\Eq{Sn} by writing the flavor epsilon tensors in the definition
of $\De$ in terms of sums of Kronecker deltas.
$K_{\rm eff}$ is not the unique term invariant under all symmetries,
but this is sufficient
to show that the symmetries do not force the effective \Kahler
term to be independent of $X$.

It is straightforward to repeat the above discussion for the case of a
strong $SU(2)$ gauge group.
The superpotential has the form
\beq
W = Q^T M Q
\eeq
where $M^T = -M$.
The theory with $F$ favors ($2F$ fundamentals) has
flavor group is $SU(2F)$.
Denoting the flavor gauge superfield by $X$, we can again define
$S$ as in \Eq{Sdefn},
which transforms as shown 
with $U$ a complexified $SU(2F)$ transformation.
We therefore have the invariants
\beq
\tr(S^n) = \tr[(M^\dagger M)^n]
- 2n \tr[X (M^\dagger M)^n]
+ O(X^2).
\eeq
It is straightforward to write $X$-dependent
\Kahler terms that are
conformally invariant as well
as flavor gauge invariant.

\end{document}